\documentclass[pss]{wiley2sp} 
\usepackage{amsmath}

\def\ket#1{\left |#1\right\rangle }

\def\expect#1{\left\langle#1 \right \rangle}

\def\w{\omega}

\usepackage{color}

\begin{document}

\title{Spin noise in a quantum dot ensemble: from a quantum mechanical to a 
semi-classical description}

\titlerunning{Spin noise in quantum dot ensembles}

\author{%
  Johannes Hackmann\textsuperscript{\Ast,\textsf{\bfseries 1}},
  Dmitry S.\ Smirnov\textsuperscript{\textsf{\bfseries 2}},
  Mikhail M.\ Glazov\textsuperscript{\textsf{\bfseries 2,3}},
  Frithjof B.\ Anders\textsuperscript{\textsf{\bfseries 1}}}

\authorrunning{J.~Hackmann et al.}

\mail{e-mail
  \textsf{johannes.hackmann@tu-dortmund.de}, 
  Phone:   +49-231-755-7958, Fax: +49-231-755-5059}

\institute{%
  \textsuperscript{1}\,Lehrstuhl f\"ur Theoretische Physik II, Technische Universit\"at Dortmund, 44221 Dortmund, Germany\\
  \textsuperscript{2}\,Ioffe Physical Technical Institute, Russian Academy of Sciences,
Polytekhnicheskaya 26, St. Petersburg, 194021 Russia\\
 \textsuperscript{3}\, Spin Optics Laboratory, St. Petersburg State University, 1 Ulanovskaya, Peterhof, St. Petersburg 198504, Russia
 }

\received{14.3.2014} 



\abstract{%
%
%
%
\abstcol{%
Spin noise spectroscopy is a promising technique for revealing the
microscopic nature of spin dephasing processes in quantum dots. We
compare the spin-noise in an ensemble of singly charged quantum dots
calculated by two complementary approaches. The Chebyshev polynomial
expansion technique (CET) accounts for the full quantum mechanical fluctuation
of the nuclear spin bath and a semi-classical approach (SCA) is based
on the averaging the electron spin dynamics over all different static
Overhauser field configurations.
 }{%
We observe a remarkable agreement between both methods in the
high-frequency part of the spectra, while the low-frequency part is
determined by the long time fluctuations of the Overhauser field.  We
find small differences in the spectra depending on the distribution of
hyperfine couplings.
The spin-noise spectra in strong enough magnetic fields where the nuclear dynamics is quenched calculated by two complimentary approaches are in perfect agreement.
}}

%
%

%

\maketitle   

\section{Introduction}

The promising perspective of combining traditional electronics with
novel spintronics devices lead to intensive studies of the spin fluctuations
in semiconductor quantum dots (QDs)
\cite{CrookerBayerSpinNoise2010,Dahbashi2012,LiBayer2012,ZapasskiiGreilichBayer2013}.
The spin noise  technique was originally developed for the observation of
magnetic resonance in sodium atoms \cite{SpinNoise1981} and is used to
monitor the spin Faraday or Kerr rotation effect on the linearly
polarized continuous wave probe.  Successfully applied to various
semiconductors \cite{Oestreich2005,Mueller2008} this approach has the
potential to reveal the intrinsic dynamics of electron or hole spins
interacting with its environment, see Refs.~\cite{Zapasskii:13,Oestreich:rev} for recent reviews.

For the spin dynamics of a single electron confined in a semiconductor
QDs various interactions play a role. The main contribution of the Fermi
contact hyperfine interaction has been identified
\cite{HansonSpinQdotsRMP2007,FischerLoss2008} described by the central
spin model (CSM) \cite{Gaudin1976}.  Charge fluctuation of donors and
acceptors and electron-phonon interactions provide additional
relaxation mechanisms \cite{Glazov2012}.  Even though the CSM is
exactly solvable~\cite{Gaudin1976}, the explicit solution is restricted
to a finite size system of $N<50$ nuclear spins
\cite{BortzStolze2007,FaribautSchuricht2013a}.

Over the last decade, a very intuitive picture for the central spin
dynamics interacting with a spin bath has emerged. The separation of
time scales \cite{Merkulov2002} -- a fast electronic precession around
an effective nuclear magnetic field, and slow nuclear spin precessions
around the fluctuating electronic spin -- has motivated various
semiclassical approximations
\cite{Merkulov2002,KhaetskiiLoss2003,Al-Hassanieh2006,ChenBalents2007,Sinitsyn2012,Smirnov2014}
which describe very well the short-time dynamics of the central spin
polarization.
Since experiments are performed on QD ensembles
\cite{CrookerBayerSpinNoise2010,Dahbashi2012,LiBayer2012,ZapasskiiGreilichBayer2013}
an averaging over the contribution of different QD has to be
performed.

In this paper, we compare the spin-noise spectra for QD ensembles
obtained using a quantum mechanical approach based on a CET
\cite{TalEzer-Kosloff-84,Dobrovitski2003,Fehske-RMP2006,HackmannAnders2014}
and semiclassical approach to spin fluctuations in singly charged QDs
\cite{Glazov2012}.  While the original
application~\cite{Dobrovitski2003} of the CET was restricted to the
propagation of a single wave function, we have used an extension to
thermodynamic ensembles to access the high temperature limit relevant
to the experiments.

Both approaches require information on the distribution of hyperfine
couplings in the QD ensemble and the fluctuating Overhauser field
generated by the nuclear spins confined in the QD. We have used the
experimentally determined distribution of diameters of the quantum
dots \cite{Leonhard1994} to obtain the distribution functions
assuming a Gaussian or an exponential electronic wave function of the
electronic bound state of the QD.

\section{Modelling the spin dynamics in  the quantum dots}
\label{sec:model}

The spin-decoherence of a single electron spin confined in a
semiconductor QD is mainly governed by the hyperfine interaction
between the electron spin $\vec{S}$ and the surrounding nuclear spins
$\vec{I}_k$~\cite{HansonSpinQdotsRMP2007,FischerLoss2008,Merkulov2002,Smirnov2014,Testelin2009}. In an applied external magnetic field
$\vec{B}=B\vec{n}_B$ ($\vec{n}_B$ being the unit vector in field
direction, and $B=|\vec{B}|$) the Hamiltonian is given by
\begin{equation}
H = \w_L \vec{S} \cdot \vec{n}_{B} + \sum_{k=1}^{N} A_k \vec{I}_k \cdot \vec{S}
\label{eq:hamiltonian}
\end{equation}
where the Larmor frequency $\w_L=g \mu_B B$ is introduced, $g$ is the
electron $g$-factor, and we put $\hbar=1$. In
Eq.~\eqref{eq:hamiltonian} the summation is carried out over the
nuclei interacting with the electron, $A_k$ are the corresponding
hyperfine constants, $N$ is number of relevant nuclear spins.  For
simplicity, we restrict ourselves to $s=1/2$ nuclear spins. In a more
realistic model of GaAs QD one has to take into account nuclear
$s=3/2$ spin states. This would make the bath spins even more
classical and does not change the qualitative behavior of the noise
spectrum unless stress-induced quadrupolar nuclear
interactions~\cite{Sinitsyn2012} are included in the calculation.

The fluctuations of the Overhauser field $\vec{B}_N = \sum_{k=1}^{N}
A_k \vec{I}_k $ define the energy scale $1/T^* = [\sum_k A_k^2]^{1/2}$
which governs the short-time spin dynamics. At the time scale $\sim
T^*$ the nuclear fields can be treated as static. In typical GaAs QDs
$T^*\sim 1$~ns~\cite{Merkulov2002}. The spin dynamics of nuclei
becomes important at a longer time scale $\sim \sqrt{N} T^*$ (being on
the order of $1$~$\mu$s). To address the dynamics at such times one
has to solve CSM.\footnote{At much longer times one has to take into
account the dipole-dipole interactions between nuclear spins which do
not conserve the total spin and may also lead to the nuclear spin
diffusion. These processes are disregarded hereinfater.} Although it
is exactly solvable using the Bethe-ansatz approach \cite{Gaudin1976}
the explicit evaluation of the spin dynamics is only possible for
small numbers of bath spins $(N<50)$ \cite{FaribautSchuricht2013a},
therefore we resort to numerical approach, see below.

\subsection{Chebyshev polynomial expansion technique}

We have applied the  CET
\cite{TalEzer-Kosloff-84,Dobrovitski2003,Fehske-RMP2006,HackmannAnders2014}
to calculate the spin autocorrelation function and the spin noise in the CSM (\ref{eq:hamiltonian}).
The CET has originally been  proposed to propagate 
single initial state $|\psi_0 \rangle$ under the
influence of a general time-independent and finite-dimensional
Hamiltonian $H$:
\begin{equation}
|\psi(t) \rangle =  e^{-i  H t}|\psi_0\rangle 
= \sum_{n = 0}^{\infty} b_n(t)
|\phi_n \rangle .
\label{CET}
\end{equation} 
The infinite set of states $|\phi_n \rangle$
obey the Chebyshev recursion relation~\cite{Fehske-RMP2006}
\begin{equation} 
|\phi_{n+1} \rangle = 2 H' |\phi_n \rangle -
|\phi_{n-1} \rangle ,
\label{chebyshev-recursion-relation}
\end{equation} 
subject to the initial condition $|\phi_0 \rangle =
|\psi_0 \rangle$ and $|\phi_1 \rangle =  H' |\psi_0 \rangle$, with the  dimensionless Hamiltonian $H'$. The latter is defined as $H' = (H-\alpha)/\Delta E$, where $\Delta E$ is the spectral width and $\alpha$ is the 
center of the energy spectrum.
The time-dependence is included
in the  expansion coefficients $b_{n}(t) =
(2-\delta_{0,n} ) i^n e^{-i\alpha t} J_n (\Delta E t)$
containing the Bessel function $J_n(x)$, $\delta_{n,m}$ is the Kronecker $\delta$-symbol. 
Since $J_n(x) \sim (e x/2 n)^n$ for large order $n$, the Chebyshev
expansion converges quickly as $n$ exceeds $\Delta E t$. This allows
to terminate the series (\ref{CET}) after a finite number of elements
$N_C$ guaranteeing an exact result up to a well defined order.
The main limitation of the approach  stems
from the size of the Hilbert space, since each of the states
$|\phi_n\rangle$ must be constructed explicitly.

For the evaluation of the spin autocorrelation function 
\begin{equation}
\label{time:corr}
\mathcal S_\alpha(t)= \left \langle S_\alpha (t) S_\alpha (0) \right \rangle =
\sum_{i = 1}^{D} \left \langle i \right \rvert
\rho_0 \text{e}^{iHt}S_\alpha \text{e}^{-iHt} S_\alpha  \left \lvert i \right \rangle
\, ,
\end{equation}
we resort to a stochastical method. 
Here $\{ \ket{i} \}$ denotes the
complete basis set of the Hilbert space of dimension $D$, $\rho_0$ is
the density operator of the equilibrium system and $\alpha=x,y,z$ labels
the Cartesian coordinates.  It has been shown \cite{Fehske-RMP2006}
that calculation of the full trace can be replace by summing $N_s$
random states $\ket{r}$ the error scales as $1/\sqrt{N_s D}$. The parameter
$D$ grows exponentially with $N$, only a few random states are
required for an accuracy evaluation of the trace for large $N$. For
calculation of the autocorrelation function, the CET is used to
propagate the two states $\ket{r_1}=S_\alpha\ket{r}$ and $\ket{r_2}=
\rho_0\ket{r}$. The noise spectrum, 
\begin{equation}
\label{freq:corr}
\mathcal S_\alpha(\omega) = \int_{-\infty}^\infty \mathcal S_\alpha(t) \mathrm e^{\mathrm i \omega t} \ dt,
\end{equation}
is obtained by an
analytical Fourier transformation of the autocorrelation function: the
spectral information is encoded in the Chebychev polynomial and the
dependence on the Hamiltonian enters via momenta generated from two
different initial states by Chebyshev recursion.  For more technical
details see Refs.~\cite{Fehske-RMP2006,HackmannAnders2014}.

\subsection{Semi-classical approach}
It has been noted that quantum mechanical simulations of the spin
dynamics up to $N=1000$ using the TD-DMRG
\cite{Schollwoeck2011,StanekRaasUhrig2013} shows remarkably good
agreement with SCA
\cite{Merkulov2002,Sinitsyn2012,KhaetskiiLoss2003,Al-Hassanieh2006,Glazov2012}
on short time scales.  Apparently, the bath spins can be replaced by a frozen classical spin for large $N$ 
on the short time dynamics
on the time scale $T^*$. In the classical picture
\cite{Merkulov2002,Glazov2012} the electron precesses fast around a
sum of the Overhauser field and the external magnetic field, while the
individual nuclear spin $I_k$ precesses slowly around the electron
spin on a time scale given by $1/A_k \sim \sqrt{N} T^* \gg T^*$. Replacing the bath
dynamics with a static Overhauser field $\vec{\Omega}_N$ given in
units of a Lamor frequency, the electron spin fluctuations $ \delta \vec{s}(t)$
can be described by the Langevin approach applied to the Bloch
equation as follows \cite{Glazov2012}:
\begin{equation}
\frac{\partial \delta \vec{s}(t)}{\partial t}
+ \frac{\delta \vec{s}(t)}{\tau_s}
+  \delta \vec{s}(t)\times (\vec{\omega}_L + \vec{\Omega}_N)
 =\xi(t)
 \, .
\end{equation}
Here $\xi(t)$ denotes the fictitious random force field. Its
correlator does not depend neither on $\bf B$ nor on $\vec{\Omega}_N$
and is given by $\expect{\xi_\alpha(t')\xi_\beta(t)} =
\delta_{\alpha\beta}\delta(t-t')/2\tau_s$, and $\tau_s$ is an
additional electron spin-relaxation time caused by, e.g.,
electron-phonon interaction. To connect the electron spin dynamics in
a single frozen Overhauser field with the quantum mechanical
calculation, we have to average over all possible Overhauser
fields. For conduction band electrons, the distribution function
${\cal F}(\vec{\Omega}_N)$ is isotropic and approaches a Gaussian
\cite{Merkulov2002} for large $N$ whose variance is given by
$1/2[T^*]^2$.  In the absence of magnetic field the spin fluctuations
are isotropic, $\mathcal S_x(\omega) = \mathcal S_y(\omega) = \mathcal
S_z(\omega) \equiv \mathcal S(\omega)$, and it has been shown that the
spin fluctuation spectrum is given by \cite{Glazov2012}
\begin{align}
\label{eq:semi-classical}
\mathcal S(\w) = \frac{\pi}{6} &
\left\{ \Delta(\omega) +\int_0^\infty d\Omega_N  F(\Omega_N)
\right . \\
\nonumber
& \times
\left.
\left[
 \Delta(\omega+\Omega_N )  +  \Delta(\omega - \Omega_N)
\right]
\right\}
\end{align}
where $F(\Omega_N)=4\pi\Omega_N^2 {\cal F}(\vec{\Omega}_N)$, 
$\Omega_N =|\vec{\Omega}_N|$ and
\begin{equation}
\Delta(\omega) = \frac{1}{\pi} \frac{\tau_s}{1 +(\omega\tau_s)^2}
\, .
\end{equation}
In the realistic limit of very slow electron spin relaxation, $\tau_s \gg T^*$, $\Delta(\omega)$ 
can be replaced by a $\delta$-function and the
integral in (\ref{eq:semi-classical}) can be solved analytically: we
recover the Fourier transformation of spin-decay function function
derived by Merkulov et al.\ \cite{Merkulov2002,HackmannAnders2014,Glazov2013:rev}
using a Gaussian distribution function ${\cal F}(\vec{\Omega}_N)$.

\subsection{Distribution function of the hyperfine couplings}

The coupling constants $A_k$ entering the CSM are proportional to the
square of the absolute value of the electron wave function at the
$k$-th nucleus \cite{HansonSpinQdotsRMP2007,FischerLoss2008}.  The
envelope of electron wave function depends on the details of the
confinement potential and band parameters of the system under study. Ignoring the microscopic details we 
use the generic form \cite{CoishLoss2004}
\begin{align}
  \psi (\left\lvert \vec{r} \right\rvert) &\propto 
  \exp{\left[-\frac12\left(\frac{|\vec{r}|}{L_0}\right)^m\right]},
\end{align}
where $L_0$ is the characteristic length scale of the QD.  For $m=2$,
$ \psi(r)$ is a Gaussian, and it takes the form of hydrogen $s$-state
for $m=1$.  Assuming a spherical shape of a $d$-dimensional quantum
dot, we find the probability distribution $P(A)$ of the hyperfine
coupling constants \cite{HackmannAnders2014},
\begin{align}
\label{eq:p-a}
  P(A) &= \frac{d}{m} \frac{1}{r_0^d \cdot A} 
  \left[ \text{ln}\left( \frac{A_{\text{max}}}{A} \right) \right]^{\frac{d}{m}-1},
\end{align}
where $A_{\text{max}}$ is the largest coupling constant in the center
of the quantum dot.  The smallest coupling constant is determined by
the cutoff radius $R$ regularising the distribution $P(A)$ and
entering the ratio $r_0 =R/L_0$.  By calculating the average
$\overline{A^2} = \int dA P(A)A^2$, we find that the characteristic
time scale $T^*(L_0)$ is proportional to $L_0^{d/2}$ and independent
of cutoff $R$.  The spin noise spectra are measured, as a rule, for QD
ensembles. To describe this case, one also has to take into account
the spread of quantum dots sizes. As a simplest possible model we
follow Ref.~\cite{Leonhard1994} and assume that the distribution of QD
radii in quantum dot ensembles can be approximated by a Gaussian with
a standard deviation $\sigma_r$ given by $\sigma_r/L_0 \approx 0.15$.

In the full quantum mechanical approach we are averaging the
spin-noise spectrum of a single CET calculation
\cite{HackmannAnders2014} over typically 100 different configurations
of randomly generated sets $\{A_k\}$ drawn from $P(A)$.  For
performing the ensemble average over many quantum dots, we use a
Gaussian distribution for the QD radii \cite{Leonhard1994} and the
scaling $T^*\propto L_0^{d/2}$ to randomly assign a characteristic
time scale $T^*(L)$ to each individual generated set $\{A_k\}$ by
proper normalization.

For the semi-classical approach \cite{Glazov2012} we assume a Gaussian
distribution of the Overhauser fields in a single quantum dot
characterised by $T^*$ as stated in Ref.~\cite{Merkulov2002} and
average this Gaussian over the same distribution of QD radii as in the
full quantum mechanical approach to obtain the nuclear spin distribution function $F(\Omega_N)$.

\section{Comparison of the two approaches}

\begin{figure} [htbp] 
\centering

\includegraphics[width=80mm, height=40mm]{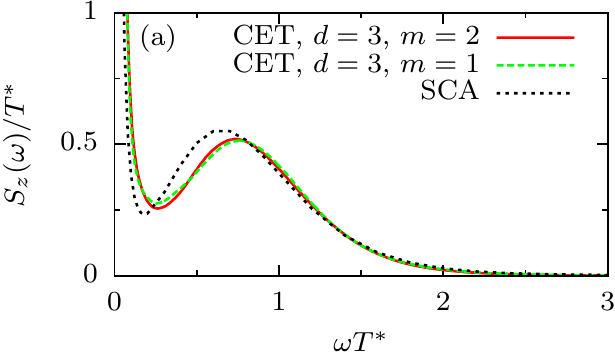}

\includegraphics[width=80mm, height=40mm]{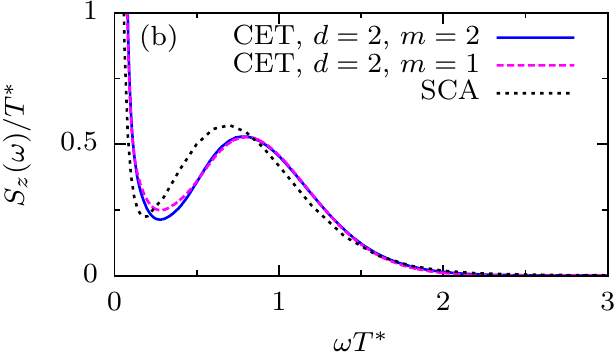}

\includegraphics[width=80mm, height=40mm]{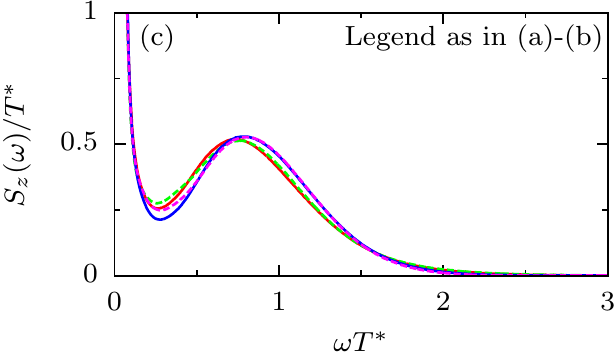}

 \caption{Analysis of the spin-noise spectrum $S_z(\omega)$ at zero external magnetic field.
Comparison of $S_z(\w)$ between the semiclassical
approach (SCA) and  fully quantum mechanical CET
for two different envelope functions, $m=1,2$, and
 $d=3$ (a) and $d=2$ (b).
Panel (c) combines the CET results of (a) and (b).
For all CET calulations
the ratio between the largest and smallest coupling constant
has been set to $A_{\text{max}} / A_{\text{min}} = e^{9/4}$ and
$S_z(\w)$ has been averaged over  $200$ individual QDs.
For each QD, we included $N=18$ nuclear spins
each coupled to the electron spin with a random
coupling constant $A_k$ generated from $P(A)$
and the distribution of  characteristic timescales.
For the SCA we set $\tau_s = 50 T^{*}$. }

  \label{fig:1}
\end{figure}
%

Now we present a comparison of the spin-noise spectrum obtained
from the fully quantum mechanical CET and SCA.

\subsection{Zero external magnetic field}

We begin with the results at $\bf B=0$.  All approaches fulfil the spectral sum
rule: the integrated spin-noise spectrum must be $1/4$, determined by
the value of the autocorrelation function at $t=0$
\cite{HackmannAnders2014}.  We have used the time scale $T^*=\langle
T^*(L_0)\rangle$ as inverse unit of energy (frequency) in all of our plots 
where we present the spin noise spectra averaged over the distribution of QD radii. 
This averaging is denoted as $\langle \cdots\rangle$ in what follows.

Since the characteristic time scale $T^*\propto L_0^{d/2}$ depends on
the QD dimension $d$, we present our results for a Gaussian ($m=2$)
and exponential envelope function ($m=1$) for $d=3$ in Fig.\
\ref{fig:1}(a) and for $d=2$ in Fig.\ \ref{fig:1}(b).  In order to
compare the finite size CET calculations for the different electronic
envelope functions, we determined the cutoff radius such that the
ratio between the largest and the smallest hyperfine coupling in the
simulation always remains at $A_{\text{max}} / A_{\text{min}} =
e^{9/4}\approx 9.4877$.

In the SCA each individual QD is characterised by a Gaussian
distribution of Overhauser fields whose width is determined by
$T^*(L_0)$.  Consequently, the SCA results are only dependent of the
dimension $d$ and the distribution of QD radii. We have obtained the
distribution ${\cal F}(\vec{\Omega}_N)$ entering Eq.\
(\ref{eq:semi-classical}) by averaging the radius dependent Gaussian
over the distribution function of the radii, i.\ e.\ ${\cal
F}(\vec{\Omega}_N) =\langle {\cal F}[\vec{\Omega}_N,T^*(L_0)]\rangle$.

Figure \ref{fig:1}(a) and (b) clearly show that the high-energy tails
of the CET spin-noise spectrum $\mathcal S(\w)$ are independent of the detailed
shape of the envelope function and perfectly agrees with the SCA
results. Only for low frequencies $\w\ll 1/T^*$ deviations of the two
approaches are observed. Those difference are related to the different
treatment of the Overhauser field: While the SCA neglects the
fluctuation of Overhauser field and performs the limit $N\to\infty$,
the quantum mechanical CET includes the full dynamics of the small
finite size nuclear spin. The slow precession of the individual
nuclear spins yields a shift of the conserved spectral weight below
the maximum $\w T^*\approx 1/\sqrt{2}$ to lower frequencies in the
CET.  Furthermore, the slight differences in the noise spectra of
$m=1$ and $m=2$ are related to the differences in the distribution
function $P(A)$ given by Eq.\ (\ref{eq:p-a}).

In figure \ref{fig:1}(c) we have combined the CET results of Fig.\
\ref{fig:1}(a) and (b) for both dimensions. 
The overall qualitative
agreement is remarkable.  The high frequency tails, however, are
clearly dependent on the dimension which can be traced back to the
different scaling of $T^*(L_0)\propto L_0^{d/2}$.  
We omitted the SCA
results in Fig.\ \ref{fig:1}(c) since the differences between the SCA data in 2D and 3D 
are similar to those of the CET and are also related to the scaling
of $T^*(L_0)$.
The coincidence of
$S(\w)$ for $\w\to 0$ is caused by the finite number of Chebyshev
polynomials entering the CET approach.
Detailed description of low frequency spin noise spectra is beyond the scope of the present paper~\cite{HackmannAnders2014}.

\subsection{Finite external magnetic field}
\begin{figure} [t] 
\centering

\includegraphics[width=80mm]{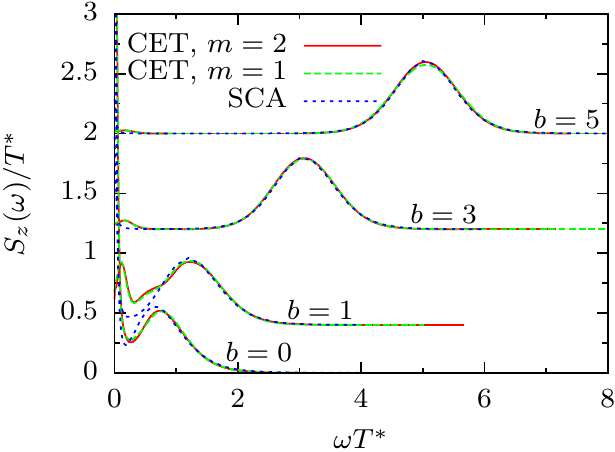}

 \caption{The spin correlation function $S_z(\omega)$ is shown for a
varying external magnetic field applied along the $x$-direction, whose
strength is given by $b = \omega_L T^*$.  For clarity an offset
proportional to $b$ has been added to the individual curves. All shown
results are based on $d=3$ dimensional QDs and the ensemble average
has been applied as described in the subset of Fig.\ \ref{fig:1}.
For the SCA we set $\tau_s = 50 T^{*}$. }

  \label{fig:2}
\end{figure}

Let us turn to the evolution of the spin-noise spectrum in a
finite magnetic field applied in $x$-direction. 
In the presence of the magnetic field, the fluctuation 
spectra of transverse, $\mathcal S_z(\omega)$, $\mathcal S_y(\omega),$ 
and longitudinal, $\mathcal S_x(\omega),$ components become 
different~\cite{Glazov2012}. In what follows we focus on the case 
of Voigh geometry and address the spin $z$-component noise 
spectrum, $\mathcal S_z(\omega)$.
Since according to Fig.\ \ref{fig:1}(c) there are only very subtle
differences between the different dimensions, we restrict ourselves to
$d=3$. We present a comparison of the SCA and the CET for three
different dimensionless magnetic fields $b = \omega_L T^*=1,3,5$ in
Fig.\ \ref{fig:2}.  For completeness, we have added the data of Fig.\
\ref{fig:1}(a), i.\ e., the curves for $b=0$.

For $b=1$, the CET peak has almost approached the SCA curve. 
The only deviations at small frequencies are related to the spectra weight located
in the $\w=0$ peak. Namely, the CET faster shifts the weight of the
non-decaying fraction of the autocorrelation function, originally
described by the $\Delta(\w)$-peak at $b=0$, to higher frequencies than
the SCA.  Futhermore, the application of the external magnetic field
suppresses the quantum-fluctuation of the nuclear spin bath the stronger, the higher 
field strength $b$.  At $b=3$, hardly any difference are
observable and for $b=5$ we found perfect agreement between the CET
and the SCA.
The peak position in the spin-noise spectra is given by 
$\omega^* = \sqrt{\omega_L^2 + (T^*)^{-2}/2}$ 
and approaches the electron Larmor frequency $\omega_L$
only at large magnetic field, $\omega_L \gg 1/T^*$ \cite{HackmannAnders2014}.

\begin{figure} [t] 
\centering

\includegraphics[width=75mm]{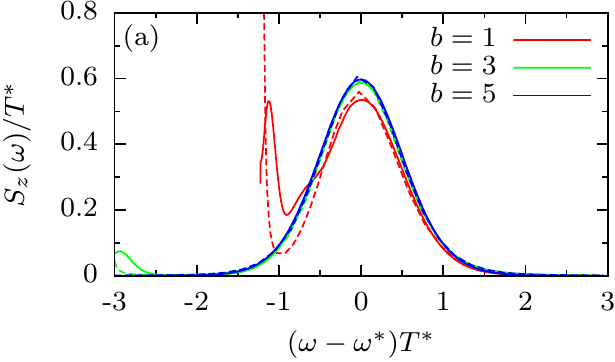}

\includegraphics[width=75mm]{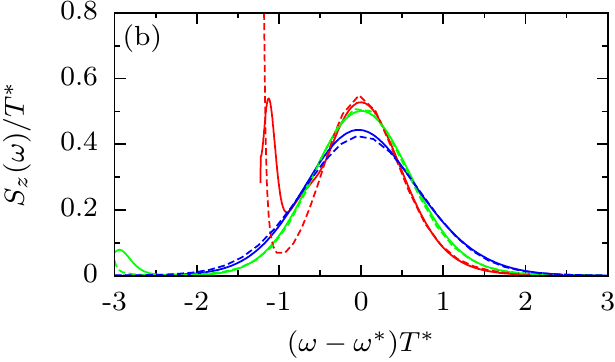}

\includegraphics[width=75mm]{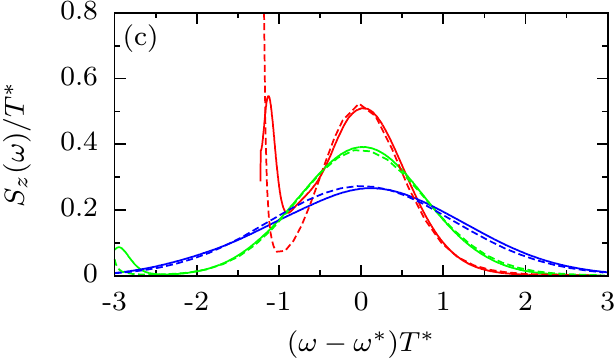}

 \caption{
 The spin correlation function $S_z(\omega)$ 
for three different a finite magnetic fields $b=1,3,5$ 
applied along the $x$-axis and 
different distributions of $g$ factors
plotted  as function of  the shifted frequency $(\w-\w^*)T^*$:
(a) the same data as Fig.\ \ref{fig:2} for a fixed $g=g_0$,
 Gaussian distributions of $g$-factors with  (b) $\Delta g/g_0=0.1$
and (c)  $\Delta g/g_0=0.2$.
The CET data  are depicted as solid line, the SCA curves as dashed line
of the same color.
For the SCA we set $\tau_s = 50 T^{*}$. }

 \label{fig:3}
\end{figure}

So far, we have not taken into account the variation of the electronic $g$-factors in a QD ensemble. 
Owing to the size quantization effects, the the $g$-factor tensors of electron and holes in QDs show not only  considerable derivations from isotropy 
but additionally vary with the size and shape of the 
QD \cite{Schwan2011}.
Our aim here is to demonstrate qualitatively the effect of $g$-factor spread, hence, 
we are not trying to present a realistic modelling 
for a specific experiment. Hereafter consider 
the  Voight geometry for the applied magnetic field  only with magnetic field directed along one of the main axis of $g$-factor tensor, and a Gaussian distribution function 
$$
P(g)= \frac{1}{\Delta g \sqrt{2\pi}}\exp\left[-\frac{(g-g_0)^2}{2(\Delta g)^2}\right],
$$
onto the spin-noise spectrum, 
where $g_0$ is the average $g_{xx}$ component of $g$-factor tensor, $\Delta g$ is the width of the distribution.

Since the differences in $S_z(\w)$ are subtle and vanish for large magnetic fields, 
we restrict our discussion on the impact of a $g$-factor distribution  to the case 
$d=3$ and $m=2$. 
The results for two different values
values  $\Delta g/g=0.1,0.2$  and three different magnetic fields $b=1,3,5$
are depicted in Fig.\ \ref{fig:3}. For completeness we added the data
of Fig.\ \ref{fig:2}(c) as panel (a), corresponding to $\Delta g=0$. The CET results are 
plotted as solid line, while the SCA data have been added 
as dashed line in the same color for the same magnetic field.
By plotting in Fig.~\ref{fig:3} the data as function of $(\w-\w^*)T^*$, we clearly show that the  finite frequency maxima
are located at the analytically predicted effective Larmor frequency $\w^*$
which approaches  the bare Larmor frequency $\w_L$ only at very large $b$-fields. 
For small fields, the spectral contribution near $\w\approx 0$ remains  visible.
Upon increasing the spread of the $g$-factors, the broadening of the spin-noise peak at $\w^*$ is increasing
with increasing magnetic field as can be seen in  Fig.\ \ref{fig:3}(b) and \ref{fig:3}(c). Again,
the CET and the SCA agree perfectly at large magnetic fields.

\section{Conclusion}

We have presented a comparison of a semiclassical approach and fully
quantum mechanical calculation of the spin-noise spectra of quantum
dot ensembles. While the CET approach is limited to small bath sizes
but treats the quantum fluctuations exactly, the SCA includes the
correct limit $N\to \infty$ but neglects the nuclear spin dynamics.

We find a perfect agreement between both approaches for the
high-frequency parts of the spin-noise spectra: the SCA and the CET
predict the same short-time dynamics of the spin autocorrelation
function. Since the CET includes the full quantum fluctuations of the
nuclear spin bath the low frequency spectrum differs between the
methods and is sensitive to the distribution function of the hyperfine
coupling constants. Application of an external magnetic field suppresses
quantum fluctuations and spin-noise spectra agree remarkably over the
whole spectral range between both methods.

\begin{acknowledgement}

We acknowledge fruitful discussions with M.~Bayer, A.\ Greilich, E.L.~Ivchenko, D.\ Stanek, G.~Uhrig, and D.~Yakovlev. Partial support from RFBR, Russian Ministry of Education and Science (Contract No. 11.G34.31.0067 with SPbSU and leading scientist A. V. Kavokin) and Dynasty Foundation is acknowledged.

\end{acknowledgement}

%
\bibliographystyle{pss}


\bibliography{spin-noise-comparison}

\end{document}